\documentclass[a4paper,11pt]{article}
\pdfoutput=1 
\usepackage{jheppub} 
\usepackage[T1]{fontenc} 

\newcommand{\be}{\begin{equation}}
\newcommand{\ee}{\end{equation}}
\newcommand{\bea}{\begin{eqnarray}}
\newcommand{\eea}{\end{eqnarray}}
\newcommand{\bpm}{\begin{pmatrix}} 
\newcommand{\epm}{\end{pmatrix}}
\newcommand{\bl}{\begin{align}} 
\newcommand{\el}{\end{align}}
\let\no\nonumber 
\def\noi{\no \\} 
\def\({\left(} 
\def\){\right)} 
\def\[{\left[} 
\def\]{\right]} 
\def\={&=}
\def\^{\wedge}
\def\a{\alpha}
\def\b{\beta}

\def\d{\delta}
\def\e{\epsilon}

\def\k{\kappa}
\def\th{\theta}

\def\l{\lambda}
\def\L{\Lambda}

\def\o{\omega}
\def\O{\Omega}
\def\na{\nabla} 
\def\p{\partial} 

\def\s{\sigma}
\def\S{\Sigma}

\def\w{\omega}
\def\mL{\mathcal{L}} 

\title{\boldmath Fluid Dynamics and Entropic Gravity}

\author{Ian Nagle,\note{For the International Moscow School of Physics / ITEP Winter School 2016. Proceedings to be published in \it{Physics of Atomic Nuclei}.}}
\affiliation{Petersburg Nuclear Physics Institute,\\Gatchina 188300, St. Petersburg, Russia}

\emailAdd{ianagle@gmail.com}

\abstract{A new entropic gravity inspired derivation of general relativity from thermodynamics is presented. This generalizes, within Einstein gravity, the ``Thermodynamics of Spacetime'' approach by T. Jacobson, which relies on the Raychaudhuri focusing equation. Here the rest of the first law of thermodynamics is incorporated by using the Damour-Navier-Stokes equation, known from the membrane paradigm for describing fluid dynamics on the horizon.}

\begin{document} 
\maketitle
\flushbottom
\section{Introduction}
\label{intro}

The general goal of this short paper is to present a new entropic gravity inspired derivation of general relativity from thermodynamics. The plan along this route will be to briefly review some background material before jumping into a new, but short, calculation. Topics will be the the geodesic ``focusing'' equation, known as the Raychaudhuri equation, 
and its use in the derivation of general relativity from the Clasius relation by Jacobson \cite{jacobson1}. The Damour-Navier-Stokes equation is introduced, and then I indicate how it too leads to gravity.

\section{Notation}
In order to briefly fix notation: I will denote an induced metric via the pullback \begin{equation} h_{ab} = g_{\alpha \beta} e^\alpha_a e^\beta_b . \end{equation} The extrinsic curvature is written as \begin{equation} K_{a b} = \nabla_\beta n_\alpha e^\alpha_a e^\beta_b = \mathcal{L}_n h_{ab}  \end{equation} and can be expressed as the lie derivative along a normal to the hypersurface. Another indispensable quantity will be the pullback of the Ricci tensor into the Gauss-Codazzi equations. This takes the form \be R_{\a \b} e^\a_a e^\b_b = \na_b K^b_a - \p_a K . \ee I define perturbations in terms of a directional covariant derivative \be  \d s^a \equiv \na_x s^a .\ee Also important is the identity \be \mL_x (B s_a) = \na_x (B s_a) , \ee which holds so long as $s^a$ is proportional to $x^a$. Throughout this paper we will use a dual-null foliation of the near-horizon Rindler metric. Coordinates are $(u,v,\theta,\phi)$, with the principle null vector to the future horizon defined as $n_a = \a \bf{d} u$, auxiliary null vector $k^a$, and each satisfying $n^a n_a = k^a k_a = 0$ and $n^a k_a = -1$. 

\section{Raychaudhuri ``Focusing'' Equation}
\label{raych}
In general a congruence of geodesics can undergo various types of deformation, which are described in terms of the extrinsic curvature of a surface through which the geodesics pass. These are, respectively: expansion, given by the trace of the extrinsic curvature, the symmetric trace-free component known as shear, and the torsion, which vanishes in general relativity but is included in this section for completeness.
\begin{align}
\theta &= K = \na_a n^a \noi
\s_{ab} &= \frac{1}{2}K_{(ab)} - \frac{1}{2}h_{ab} \theta \noi
\o_{ab} &= \frac{1}{2}K_{[ab]}
\end{align}
The flow of the expansion scalar along a normal to a null hypersurface is known as the null Raychaudhuri equation: 
\be
\mL_n (\th dA)= \(\k \th -\frac{\theta^2}{2} - \s^2 + \w^2 - R_{ab} n^a n^b \) dA. 
\ee 
The fractional change in area of a surface element on the horizon can be seen using Raychaudhuri's equation to evolve as \be \theta = \frac{\mL_n(dA)}{dA} . \ee
%

\section{Clasius Relation to Gravity} \label{thermtograv}
A brief outline of the derivation appearing in \cite{jacobson1} follows. The idea here is to begin with the thermodynamic Clausius relation, and then through examining the ways in which the energy, entropy and temperature are defined, to recover the Raychaudhuri equation, and general relativity. Assuming the proportionality of entropy to area, which has its origin in the Bekenstein-Hawking entropy $dS = \e dA $, will be a crucial step.  

Starting from $\d Q = T dS$, let us express the heat flux on the left-hand side in terms of a boost vector along a timelike curve. This can be written as 
\be
\d Q = \int T_{ab} \chi^a d\S^b  
\ee
where $\chi^a = - \k \l n^a$ and $d\S^b=n^a d\l dA$. 

In order to progress with the other side of the Clausius relation, first invoke the Bekenstein-Hawking proportionality of entropy with surface area. 
\be
\d A = \int \theta d\l dA . 
\ee
We now use the Raychaudhuri equation to get an expression for $\theta$. Since the null horizon is stationary, the Raychaudhuri equation can be integrated to give $\theta = - \l R_{ab} n^a n^b$. Inserting this expression for $\theta$ we are left only in need of a way of describing the temperature appearing in the Clausius relation. Since this discussion is in Rindler space, it is natural to identify $T$ with the Unruh-Hawking temperature of an accelerated observer in flat spacetime, $T=\hbar \k / 2 \pi$. 

When these results are equated, after a few lines one arrives at the Einstein equations in full generality 
\be 
R_{ab} - \frac{R}{2}g_{ab} + \L g_{ab} = \frac{2\pi}{\hbar \e} T_{ab}. 
\ee

\section{``Flow'' equation = Damour-Navier-Stokes}
Recall the Raychaudhuri equation in section \ref{raych}. It can be viewed as expanding out the null-null projection of the Ricci tensor in terms of geodesic shear and expansion. Alternatively, one could instead consider expanding the null-tangential projection by using the Gauss-Codazzi equation \cite{damour1} \cite{damour2} \cite{pricethorne}. When the resulting terms are reorganized into shear and expansion components, an equation with the same form as the Navier-Stokes equations arises. This describes the tangential flow of deformations along a horizon. 
\be \label{dns}
\mL_n (\O_A dA) = \( \na_A (\k + \theta) - \na_a \(\s^a_A + \frac{1}{2}h^a_A \theta \) + R_{ab} n^a h^b_A \) dA.
\ee 
Here the role of momentum density is played by $\O_A = k_b \na_A n^b$, which is known as the Hajicek 1-form. An important conceptual difference between the Damour-Navier-Stokes equation and ordinary fluid dynamics is that due to the role of the equivalence principle in general relativity one can always shift away any localized mass density. The Damour-Navier-Stokes equation is inherently nonlocal, while the Navier-Stokes is not.

\section{Fluid Dynamics to Gravity}
Rather than the Clausius equation as in section \ref{thermtograv} consider beginning with a work of the form 
\be 
\d E = \d W = \O \d J .
\ee 
In this case we can define the momentum flux on the horizon in terms of an angular momentum Killing vector so that 
\be 
\d E = \int T_{ab} \phi^a d\S^b = \O \int T_{ab} n^a h^b_A d\l dA .
\ee 
The total angular momentum of a spacetime, when it can be defined, is given by the Komar angular momentum. This can be pulled back to the horizon and expressed succinctly using the Hajicek 1-form as 
\be
J = \rho \O \int \O_\phi dA ,
\ee
using the shorthand expression $\O_\phi \equiv \O_A \phi^A$. The Hajicek 1-form can be thought of as the ``surface density of linear momentum'' pulled back to the horizon. The dimensionful constant $\rho$ is used to fix factors involving Newton's constant. From this the work can then be written as 
\be
\d W = \O \d J = \rho \O \int \mL_n(\O_\phi dA) 
\ee
by using the identity $\d J \equiv \na_n J = \mL_n J$ to express the work in terms of the Damour-Navier-Stokes evolution equation. By appealing to the vanishing of shear and expansion on the horizon, we may simplify the Damour-Navier-Stokes equation \ref{dns} so that the work becomes
\be
\d W = \rho \O \int \mL_n (\O_\phi dA) = \rho \O \int R_{ab} n^a h^b_A dA . 
\ee 
Alternatively, if Carter-Thorne coordinates of the form $v' = v+f(x)$ are used \cite{pricethorne}, then vanishing shear and expansion is not required. Combining both expressions, 
\begin{align}
\d E &= \d W \noi
T_{ab} n^a h^b_A &= \rho R_{ab} n^a h^b_A \noi
G_{ab} + \L g_{ab} &= 8 \pi G T_{ab} 
\end{align}
In this section neither the Bekenstein-Hawking entropy or the Unruh temperature were assumed. 

\section{Conclusions}
Here a new entropic gravity inspired derivation of general relativity from thermodynamics has been developed. The net effect of considering both derivations together is that one is effectively beginning with the nonrelativistic laws of thermodynamics, specifying the conservation of Killing quantities so that global momenta and energy are well-defined, and then from this evolving the system backwards in Rindler space in order to obtain, or reexpress, general relativity in terms of thermodynamics. How far this alluring approach may evolve in the future remains to be seen. 

\acknowledgments
It is a pleasure to acknowledge the hospitality of the Petersburg Nuclear Physics Institute theory department, and in particular L. Lipatov and V. Kim, where portions of the above material were developed. I was also a student at the University of Amsterdam during the early development of these ideas. 


\end{document}